# PAR: Petal Ant Routing Algorithm for Mobile Ad Hoc Network

Manjunath M[1] and Dr. Manjaiah D.H[2]

[1]Research Scholar, Dept. of Computer Science, Mangalore University, Mangalore, India
[2]Professor, Dept. of Computer Science, Mangalore University, Mangalore, India

## ABSTRACT

*During route discovery of mobile ad hoc network, broadcasting of route request and route reply packets are the essential operations for finding the path between two ends. In such situations, intermediate node which may or may not belongs will participate in route discovery process, update routing table and rebroadcast the route discovery packets again to its neighboring nodes. Finally optimal path is found with minimum hops. This simply upsurges overhead and deteriorates the performance of routing. The proposed Petal Ant Routing (PAR) algorithm offers a low overhead by optimizing FANT and BANT transmissions in route discover process. The algorithm is an improved version of SARA and has features extracted from petal routing. The algorithm is simulated on NS2, compared with ACO frame work called SARA and classical routing protocols such as AODV and AOMDV. The simulation results shows that PAR further reduces overhead by eliminating redundant FANT transmission compared to other routing algorithm.*

## 1. INTRODUCTION

A mobile ad hoc network (MANET) [1] is a network consisting of a set of mobile nodes with no centralized administration. MANET is self-configuring, self-organizing and self-maintaining. MANET may have dynamic topology. Mobile ad hoc network (MANETs) [1] [2] are special kind of infrastructure less wireless ad hoc network. In Manets, each node acts as router and a host at the same time which joins and leave network at any movement of time with high

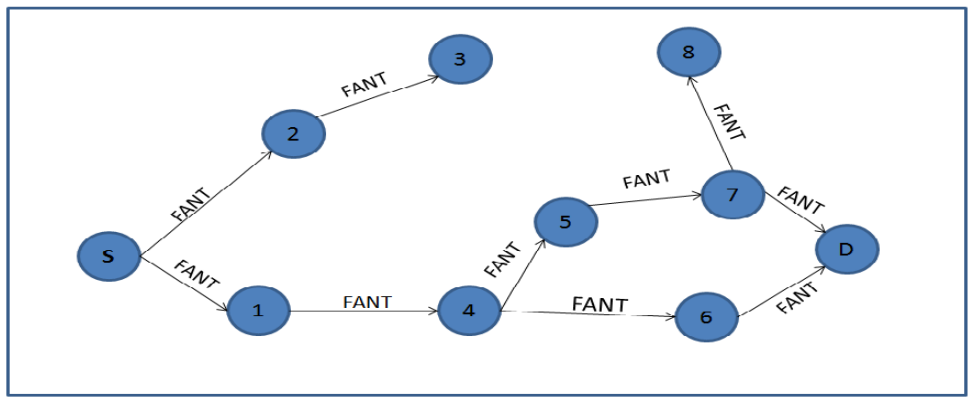

Figure 1. FANT Transmission by ANT Routing

  45

mobility [3]. Due to nodes high mobility, the topology of the network is subject to change frequently and routing for such a situation becomes difficult. The design of mobile ad hoc routing protocols is extremely challenging task because of limited bandwidth, limited power, and unpredictable radio channel behavior and node mobility [4]. The main challenges of routing protocols for Manets are to ensure that nodes are able to select an optimal path to forward the data traffic from source to the intended destination. Many routing protocols have been proposed for routing issues such as AODV [5], AOMDV[6], DSR[7], DSDV [8], TOHIP [9], S-AODV [10], S-DSDV [21],...etc, but many researchers have stated in literature that the Ant have the better potential to find an efficient and shortest path much optimal than other routing algorithm by depositing chemical substance called pheromone [2 indexing]. The researchers observed the behaviour of real ants and inspired to design a new ant routing protocols for manet such as ACO[11], ARA[12], SARA[13], HOPNET[14], ANTnet [15], Ant AODV [16], ANTALG [17], etc. In popular population-based meta-heuristic ACO algorithm, when source requires a path to destination, source broadcast special kind of packet called Forward Ant [FANT] to its neighbor nodes, which replicates and rebroadcast the FANT until it reaches destination. The destination node then destroys the FANT and reply with special packet called Backward Ant [BANT] through the intermediate nodes. Upon the reception of BANT, the source starts sending data to the destination through the shortest path. But in SARA [13], it works with the mechanism called Controlled Neighbor Broadcast [CNB], in which every node broadcast FANT to its neighbor but only one of them rebroadcast again to its neighboring node. According to author [13] [18] since FANT packet replicated by all network nodes and the network is flooded with control information will reduce its performance. Figure 1 illustrates the FANT propagation of Ant routing algorithms. As the network grows large, the large number of network nodes joins in FANT and BANT transmission which significantly increases overhead and deteriorates its performance [13]. The flooding mechanism of this FANT and BANT transmission in the network is the disadvantage and increases the time required to discovering route during route discovery [13]. The aim of all routing protocols for data transmission is to find shortest path and optimal path between end nodes, but even though a network is loaded with large number of nodes, all most all routing protocols chooses minimum hops for establishing a shortest path between source and destination and eliminates all other nodes during route discovery. So flooding FANT packets for all redundant nodes during route discovery significantly increases additional time to update routing table and increases overheads. Hence, the aim of our proposed work is to minimize FANT transmission during route discovery and to reduce overhead.

The remaining part of this research article is organized as follows. Section 2 describes the literature survey. Section 3 presents the proposed work. Section 4 presents the Simulation results and comparison. Finally conclusion, Appendix and acknowledgement is described at the end of the report.

## 2. LITERATURE SURVEY

Fernando Correia, Teresa Vazao [13] proposed an improved version of ACO framework called Simple Ant Routing Algorithm (SARA) for the mobile ad hoc network. SARA uses the concept of Controlled Neighbor broadcast (CNB) mechanism to control packet flooding during route establishment and uses deep search procedure to recover route during route repair.

Petal routing [19] is a routing algorithm for MANET. In this routing approach it merge the concept of multipath and geographic routing algorithm, where network nodes are addressed based on geographic location rather than IP address with no routing principle. All the data packets are flooded in the network but the nodes which lies inside the petal region will rebroadcast again to its neighboring nodes.

## 3. PETAL ANT ROUTING (PAR) ALGORITHM FOR MOBILE AD HOC NETWORK

In this section, we present details of PAR architecture constructs as similar to others routing algorithm. PAR algorithm is an improved version of Simple Ant Routing Algorithm (SARA)[13] and combines the few characteristics of Petal routing [19]. PAR consists of 3 phases namely Route discovers, Route maintenance and Route repair.

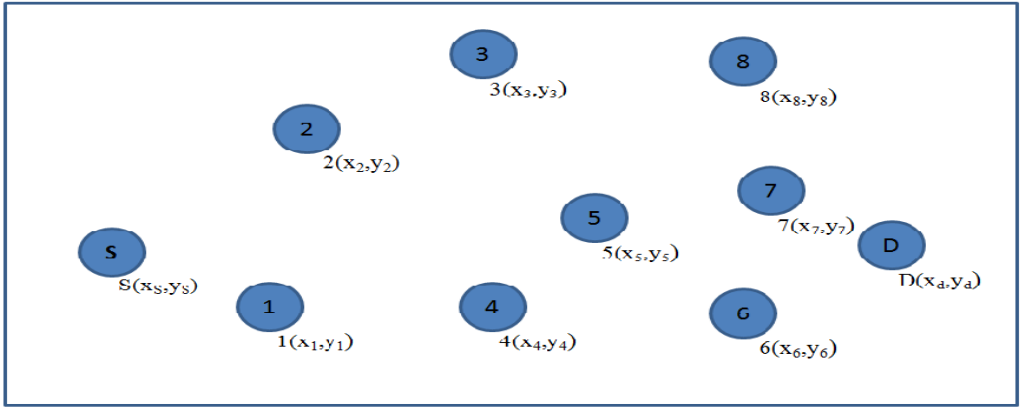

Figure 2. Network Diagram

### 3.1. Route Discovery

In the route discovery phase, PAR computes the width of the petal ($P_w$), create new routes by forwarding special packet called Petal Forward Ant [P _FANT] by source and Petal Backward Ant [B_FANT] by the destination. A P_FANT is a small packet consists of $P_w$ and unique sequence number. One key aspect of this process is to compute the petal region between end nodes and to rebroadcast the P _FANT is describes next. With this, it is possible to minimize the overhead by eliminating redundant P_FANT and P_BANT transmission during route discovery. Thus maximize the ratio of packets generation and minimize the overhead. Consider the Figure 2, Source denoted as $S(x_s, y_s)$, Destination $D(x_d, y_d)$ and the intermediate node $i(x_i, y_i)$ where i = 1,2,3,..,n. Our proposed work merges the concepts of geographic routing [20].The (x, y) coordinate of a mobile node represents (longitude, latitude) respectively. Each node is uniquely addressed inside or outside petal by geographic locations and by node id. When source (S) requires a path to destination (D), source computes the $P_w$ by following 3 steps.

Step 1: obtain nodes location dynamically and compute the distance (d) from S and D using (1)

$$d = \sqrt{(x_d - x_s)^2 + (y_d - y_s)^2} \qquad (1)$$

Step 2: Compute and obtain (h, k) using (2)

$$h = (x_s + x_d)/2, \, k = (y_s + y_d)/2 \qquad (2)$$

Step 3: Compute petal region or width of the petal ($P_w$) between S and D using (3) as shown in Figure 3.

$$P_w = \Pi ab \qquad (3)$$

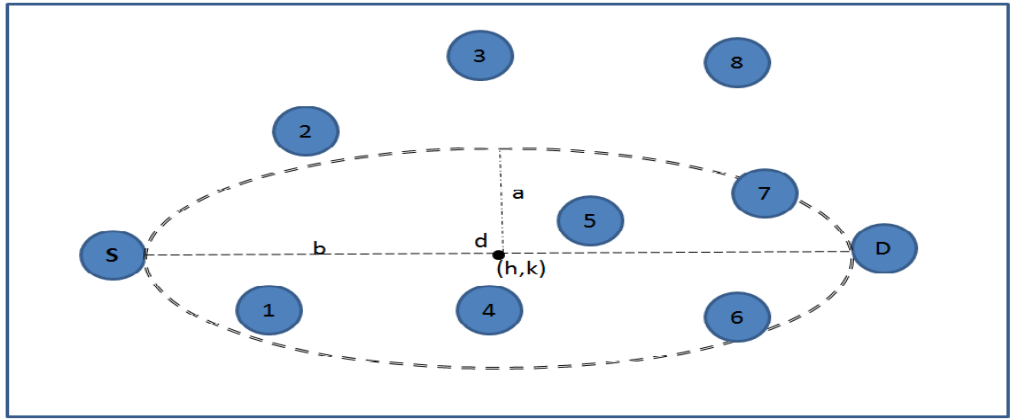

Figure 3. Petal Calculation

Once the petal region is calculated, the source (S) starts to broadcast P_FANT packet to the network. From the Figure 3, S has two adjacencies node i.e. node (1) and node (2), as the node receive the P_FANT by source node for the first time, the node verifies the $P_w$ in P_FANT and verify whether it lies inside or outside the petal region. PAR uses (4) to verify node whether it lies inside the petal or outside the petal region.

$$\frac{(x-h)^2}{a^2} + \frac{(y-k)^2}{b^2} \leq 1 \qquad (4)$$

The node which lies inside the $P_w$, will accept the P_FANT, update the pheromones value, destination address, next hop and rebroadcast the P _FANT to the next neighboring nodes. The node which doesn't lies inside the $P_w$ will discard the P_FANT and does not participate in route discovery process. This process is continued until it reaches destination (D). Once the P_FANT reaches the destination, destination extracts the information from P_BANT through the shortest path. Upon the reception of P_BANT packet, source starts transmitting data through shortest path updated by each intermediate node in the network. Thus PAR reduces overhead by eliminating the redundant FANT transmission in the network. Thereby increases the more number of packets generated by source, more number of packet received by destination and provides the better performance compare to SARA, AODV, and AOMDV.

Figure 4. PAR P_FANT Transmission

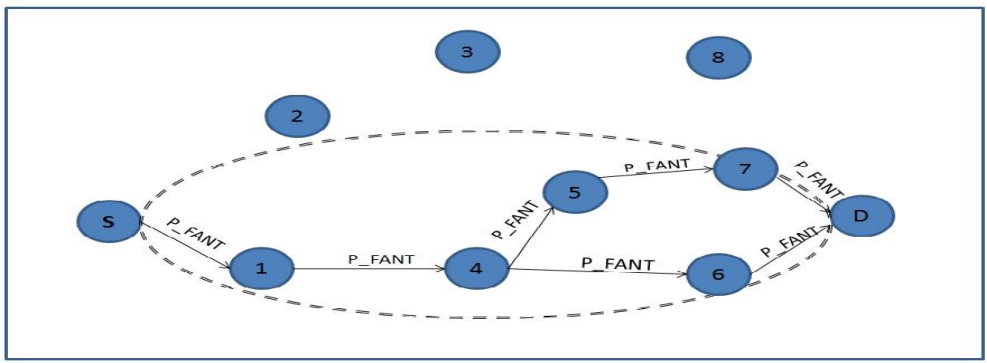

International Journal of Computer Networks & Communications (IJCNC) Vol.7, No.2, March 2015

Figure 4 and Figure 5 schematically depict the PAR route discovery process. The main improvements made on SARA to design the PAR is in receiver FANT(p) section of the algorithm. The P_FANT transmission mechanism of PAR is explained in following pseudo code.

```
OnreceivingAPacket()
if recvSaraPkt(Packet* p) then
        if (SARA _FANT) then
                Obtain node location dynamically
                Compute (d) and (h, k) using (1) and (2) respectively
                Compute width of the petal (Pw) using (3)
                Broadcast P _FANT [Pw +FANT] in the network
                 if (neighbor node = destination) then
                 |      Process routing normally
                end
                if (neighbor node lies inside Pw) then
                 |      Process routing normally
                else
                 |      Drop P_FANT
                end
        end
end
```

## 3.2. Route Maintenance

The second phase in PAR is route maintenance, which is mechanism to keep track improvements of routes and active route during communication. In ARA, there is no special packet is created for route maintenance. But in SARA, a Super FANT is created for asymmetric traffic. The PAR algorithm also updates active route while date session is running and work similar to SARA routing algorithm.

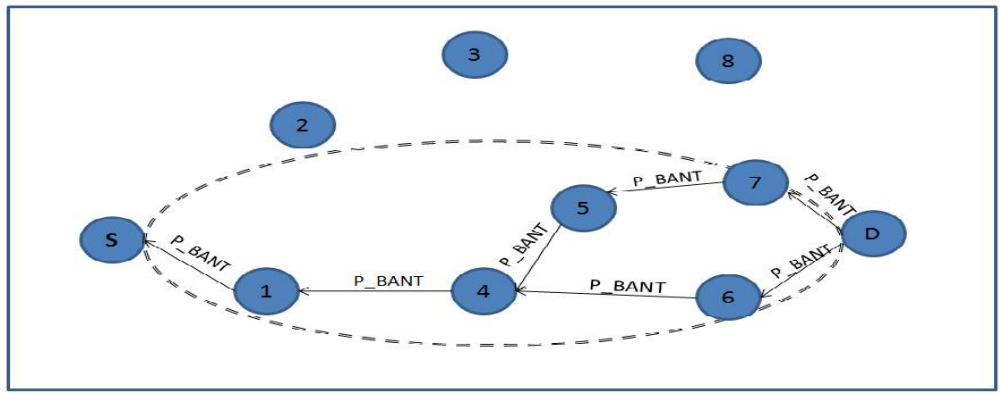

Figure 5. PAR P_BANT Transmission

International Journal of Computer Networks & Communications (IJCNC) Vol.7, No.2, March 2015

| Channel | Channel/wireless |
|---|---|
| Mobility Model | Random way point |
| Network interface | Wireless |
| NS2 version | NS2.35 |
| Interface queue | Drop Tail |
| No. of nodes | 15, 30, 50, 75, 100, 150 |
| Simulation area size | 1000 x 1000 m |
| Transmission Range | 250 m |
| Simulation Time | 160s |
| Data packet size | 512 bytes |
| Initial node energy | 100 Joule |
| Rx Power | 35.28e-3 |
| Tx Power Tx Power | 31.32e-3 |
| Idel Power | 712e-6 |
| Transport protocol | TCP |

Table 1. Simulation Parameters

### 3.3. Route Repair

The PAR initiate route repair process when broken link between two nodes is detected. Since the nodes are highly dynamic and mobile in nature, the broken link state can happen at any interval of time. This may due to node being turned off, or by limited band width or by congestion occurred, or by pheromone evaporation during data transmission. To repair route, PAR find alternative link in its routing table of the broken link. If there exists any other link between source and destination its sends the packet via this path else, if the route repair procedure fails during searching an alternative path to destination, source initiate a new route discovery process upon the reception error message.

## 4. SIMULATION EXPERIMENT SETUP USING NS2

The simulation experiment is carried under Ubuntu Linux. The proposed work and SARA of Fernando Correia et.al [NS2 version 2.31] were implemented in NS2 version 2.35 by the author's. Comparison with classical routing such as AODV and AOMDV of NS2 package is also provided. NS2 implementation of SARA code has been enhanced to reduce overhead by eliminating redundant FANT transmission during route discovery process. The simulation is

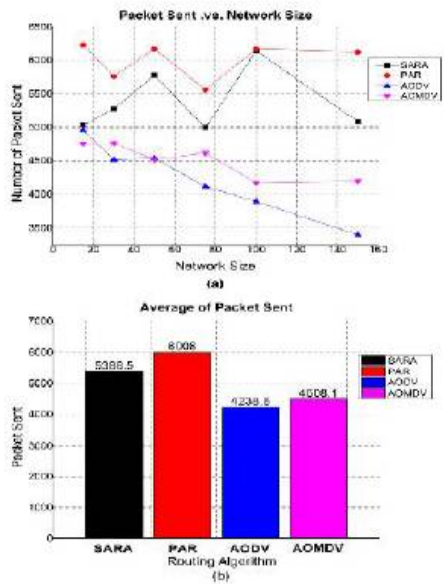 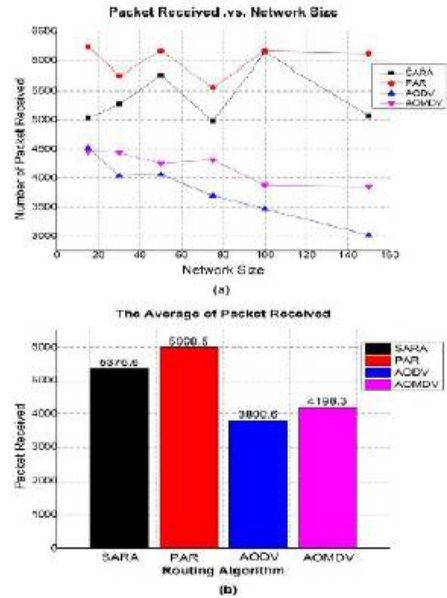

| Figure 6. Comparison of packet sent | Figure 7. Comparison of packet received |
| --- | --- |
| (a) Network Size vs. Packet generated | (a) Network Size vs. Packet Received |
| (b) The average number of packet generated | (b) The average number of packet received |

carried for two different environment (simulation environment A, simulation environment B) as described in 4.2 and 4.3.

## 4.1. Metrics considered for evaluation

The following metrics are considered to evaluate and to compare the performance of PAR, SARA, AODV and AOMDV.

*Packets sent*     - represents the total number of packets generated by all sources.
*Packets Received*     - represents the total number of packets received by all destinations.
*Packet Delivery Fraction/Ratio (PDF/PDR)* - represents the ratio of packet received by all destinations to those generated by all sources.
*End to End Delay*     - represents the average time interval taken for a packet to transmit successfully from source to destination.
*Throughput*     - represents the total number of packet delivered per unit time. it is measured in kbps.
*Overhead*     - represents the ratio between the amount of routing message generated and forwarded across the network.
*Energy Consumption* - represents the total amount of energy consumed by all the mobile nodes and measured in joules.

## 4.2. Simulation Environment A

In the first environment experiment setup, a network is loaded with moving nodes and configured with same speed. In each simulation test, nodes are generated randomly, contains only one source and the destination, move according to random way point mobility model. The simulation is carried for 160 sec and node move with a speed of 0 m/s to a maximum speed of 10 m/s. The exact values used for the number of nodes and important parameter are described at Table 1.

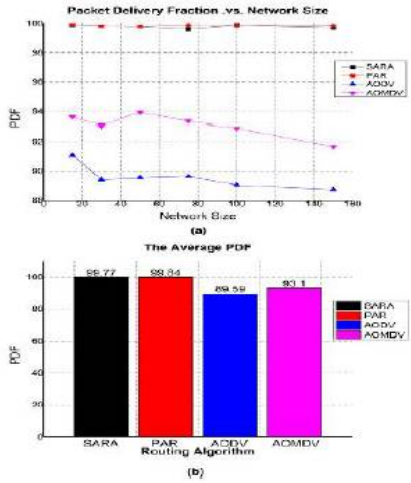
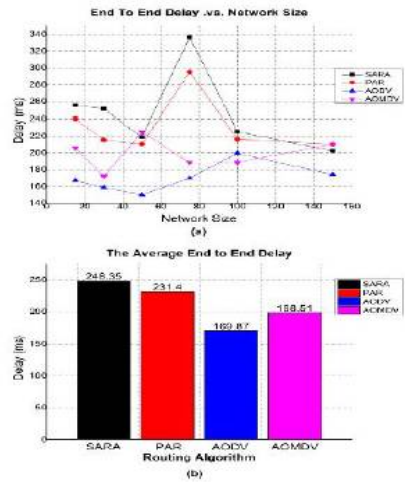

Figure 8. Comparison of Packet Deliver Ratio  
(a)Network Size vs. PDF  
(b)The average PDF

Figure 9. Comparison of End to End Delay  
(a) Network Size vs. End to End Delay  
(b) The average End to End Delay

The results of simulation are tabulates and depicted in the Table 2, Table 3, Table 4 and Table 5 in annexure. In each sets of graphs, the performance of proposed work and other algorithms such as SARA, AODV and AOMDV with respect to number of nodes is shown line graph (a) and the average performance of all routing algorithms is shown in bar chart (b).

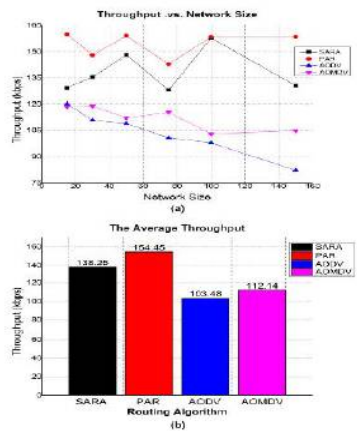
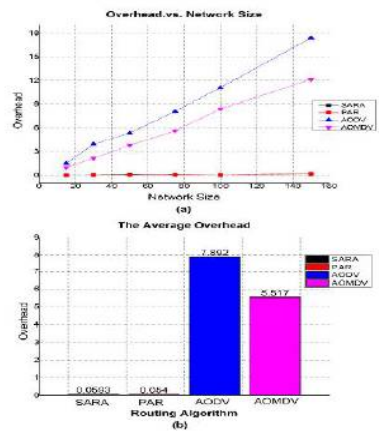

Figure 10. Comparison of Throughput  
(a)Network Size vs. Throughput  
(b)The average Throughput

Figure 11. Comparison of Overhead  
(a) Network Size vs. Overhead  
(b) The average Overhead

The graph in Figure 6(a), (b) shows the performance of four routing algorithms in terms of packet generated by the source. The proposed PAR generated more number of packets than SARA, AODV and AOMDV. From the graph of Figure 6(b), the PAR generates 11.496 % more number of packets than SARA, 41.7449 % more number of packets than AODV and 33.271 % more number of packets than AOMDV.

The graph in Figure 7(a), (b) shows the performance of four routing algorithms in terms of packets received by the destination. The proposed PAR receives the more number of packets than SARA, AODV and AOMDV. From the graph of Figure 7(b), the PAR receives 11.496 % more number of packets than SARA, 57.846 % more number of packets than AODV and 42.8411 % more number of packets than AOMDV.

The graph in Figure 8(a), (b) shows the performance of four routing algorithms in terms of Packet Delivery Fraction (PDF). The proposed PAR performance good with higher PDF than SARA, AODV and AOMDV. From the graph of Figure 8(b), the average PDF of PAR is 0.070 % more than SARA, 11.441 % more than AODV and 7.23 % more than AOMDV.

The graphs in Figure 9(a), (b) shows the performance of four routing algorithms in terms of end to end delay. The end to end delay of PAR is less SARA but higher than AODV and AOMDV. From the graph of Figure 9(b), the average end to end delay is 6.825 % less than SARA, 36.22 % more than AODV and 16.568 % more than AOMDV.

The graphs in Figure 10(a), (b) shows the performance of four routing algorithms in terms of throughput. The proposed PAR provides the better throughput when compares with SARA, AODV and AOMDV. From the graph of figure 10(b), the average throughput 11.7179 % more than SARA, 49.32 % more than AODV and 37.729 % more than AOMDV.

The graphs in Figure 11(a), (b) shows the performance of four routing algorithms in terms of overhead. The proposed PAR reduces overhead by eliminating redundant FANT transmission during route discovery when compared to SARA. From the graph of Figure 11(b), the average overhead is 8.93 % less than SARA, 99.315 % less than AODV and 99.02 % less than AOMDV.

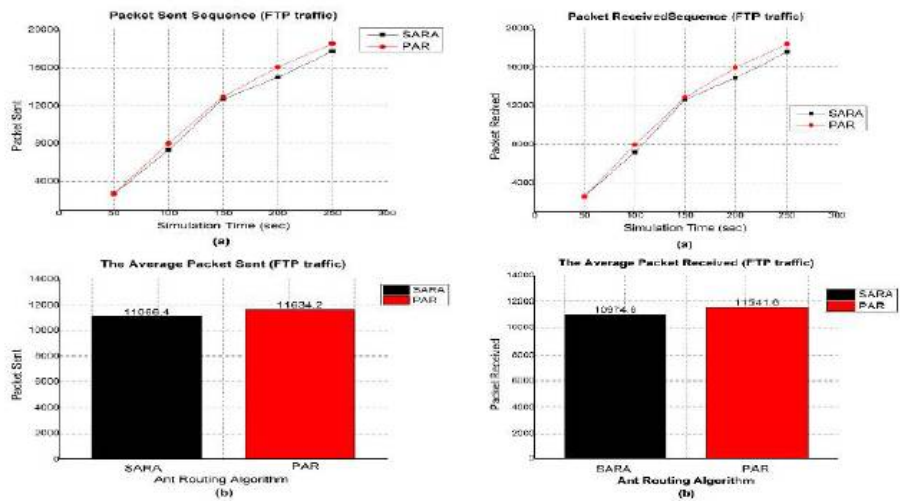

Figure 12. Comparison of Packet Generated     Figure 13. Comparison of Packet Received
(a)Network Size vs. Packet Generated     (a) Network Size vs. Packet Received
(b)The average Packet Generated     (b) The average Packet Received

International Journal of Computer Networks & Communications (IJCNC) Vol.7, No.2, March 2015

### 4.3. Simulation Environment B

In the second simulation environment experimental setup, a network is loaded with 104 wireless moving nodes in a 1000 x 1000 flat space and consists of four sources of FTP/TCP type and four destinations. All the nodes changes their location during simulation run except the destination nodes, where destination nodes were placed at centre of the scenario. The radio propagation range of each node is 200 m. Each data packet is of 1000 byte of size. The simulation is carried for 50, 100, 150, 200 and 250 seconds with a speed of 0 m/s to a maximum speed of 10 m/s. The results of simulation are tabulated and depicted in the Table 6 and Table 7 in annexure.

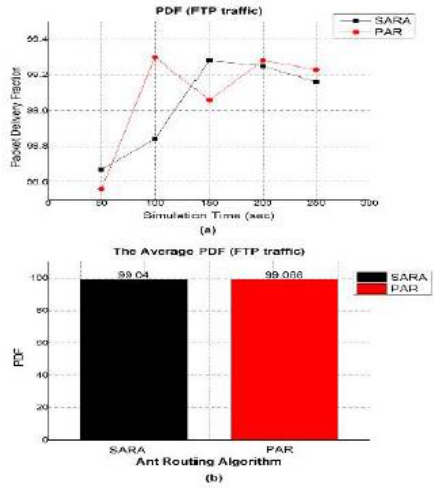 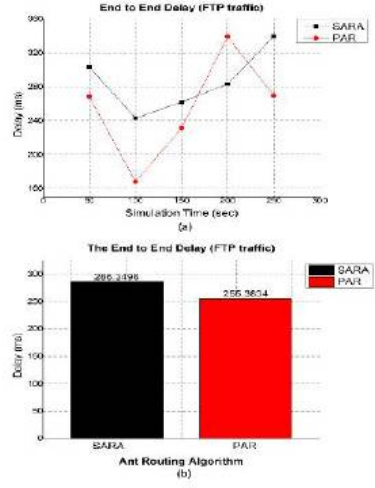

Figure 14. Comparison of Packet Deliver Fraction.  Figure 15. Comparison of End to End Delay
(a) Network Size vs. PDF                          (a) Network Size vs. End to End Delay
(b) The average PDF                                (b) The average End to End Delay

The graph in Figure 12(a), (b) shows the performance of two routing algorithms in terms of packets generated by sources with simulation time. The proposed PAR generates more number of packets than SARA. From the graph of Figure 12(b), the PAR generates 5.13 % more number of packets than SARA.

The graph in Figure 13(a), (b) shows the performance of two routing algorithms in terms of packet received by the destination. The proposed PAR receives more number of packets when compared to SARA. From the graph of Figure 13(b), the PAR receives 5.14 % more number of packets than SARA.

The graph in Figure 14(a), (b) shows the performance of two routing algorithms in terms of PDF. The proposed PAR performance is better when compared with SARA. Even though PAR generates and receives more number of packets, the percentage of Packet Delivery Fraction of PAR is better in most of cases when compared to SARA. From the graph of 14(b) the average PDF of PAR is 0.046 % more than SARA.

The graph in Figure 15(a), (b) shows the performance of two routing algorithms in terms of end delay. In most of the cases, PAR shows better performance by reducing the end to end delay. From the graph of Figure 15(b), the average end to end delay is10.814 % less than SARA. The graph in figure 16(a), (b) shows the performance of two routing algorithms in terms of throughput. As the throughput in case of SARA is little low. From the graph of Figure 17(b), the average throughput of PAR is 6.593 % more than SARA.

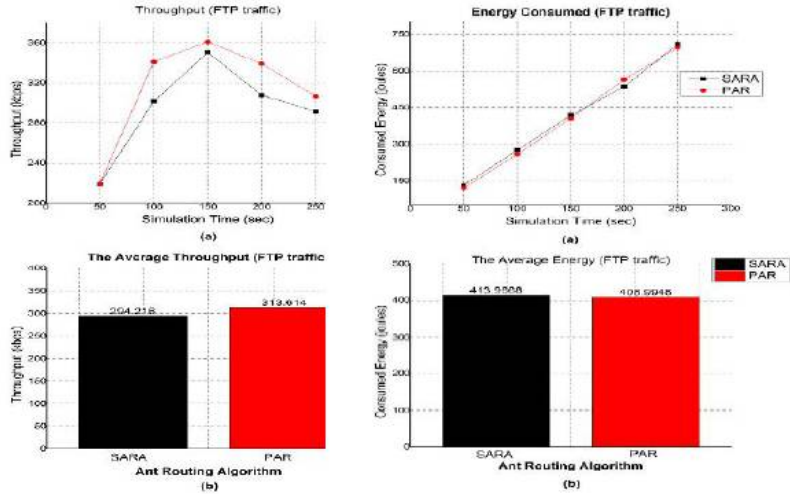

Figure 16. Comparison of Throughput
(a)Network Size vs. Throughput
(b)The average Throughput

Figure 17. Comparison of Energy Consumed
(a) Network Size vs. Energy Consumed
(b) The average Energy Consumed

The graph in Figure 17(a), (b) shows the performance of two routing algorithms in terms of energy consumption with respect to number of nodes and simulation time. The energy consumption of SARA is high when compared to PAR. From the graph of Figure 17(b) the PAR consumes 1.2010 % less amount of energy than SARA.

## 5. CONCLUSIONS

PAR (Petal Ant Routing) is an ant based routing algorithm for mobile multi-hop ad hoc network which extract few features from petal routing for computing width of the petal ($P_w$) and make P _FANT and P _BANT to propagate for establishing path between the end nodes within the petal region. The PAR algorithm has been successfully simulated using NS2. The performance of PAR has been evaluated based on different metrics and different simulation environments. The simulation results of both environment shows, the PAR performs better in terms of packet generated by the sources, packet received by the destinations, packet delivery fraction, provides good throughput and by reducing the overhead. PAR also increases network life time by reducing the end to end delay and amount of energy consumed specially in case of SARA.

## APPENDIX

The performances of routing protocols were studied with respect to different network size. We conducted the experiment tests for 15 times and only the average value is considered in each case. The experimental result shows that our proposed model performs better than SARA, AODV and AOMDV routing protocol. Table 2, Table 3, Table 4, Table 5 shows the analysis results of SARA, PAR, AODV and AOMDV. Table 6, Table 7 shows the analysis results of SARA and PAR for 104 wireless mobile nodes respectively.

Table 2. Results of SARA routing algorithm

| Network Size | Packet Sent | Packet Received | PDF | EED | Throughput (kbps) | Overhead | Consumed Energy |
|---|---|---|---|---|---|---|---|
| 15 | 5037 | 5031 | 99.88 | 255.339 | 129.38 | 0.012 | 68.6692 |
| 30 | 5277 | 5267 | 99.81 | 251.408 | 135.47 | 0.033 | 144.3165 |
| 50 | 5782 | 5769 | 99.78 | 218.957 | 148.25 | 0.066 | 227.9218 |
| 75 | 4996 | 4977 | 99.62 | 336.584 | 128.07 | 0.052 | 340.3608 |
| 100 | 6148 | 6140 | 99.87 | 225.367 | 157.82 | 0.026 | 440.9279 |
| 150 | 5091 | 5076 | 99.71 | 202.467 | 130.56 | 0.167 | 657.3948 |
| **AVG** | **5388.5** | **5376.6** | **99.77** | **248.35** | **138.25** | **0.0593** | **313.26** |

Table 3. Results of PAR routing algorithm

| Network Size | Packet Sent | Packet Received | PDF | EED | Throughput (kbps) | Overhead | Consumed Energy |
|---|---|---|---|---|---|---|---|
| 15 | 6238 | 6232 | 99.9 | 240.508 | 159.97 | 0.003 | 68.1111 |
| 30 | 5768 | 5757 | 99.81 | 215.261 | 148.01 | 0.032 | 138.04505 |
| 50 | 6176 | 6163 | 99.79 | 210.639 | 159.04 | 0.036 | 230.2723 |
| 75 | 5565 | 5557 | 99.86 | 295.162 | 142.85 | 0.69 | 325.4616 |
| 100 | 6174 | 6165 | 99.85 | 216.429 | 158.42 | 0.03 | 444.9251 |
| 150 | 6127 | 6117 | 99.84 | 210.409 | 158.42 | 0.154 | 645.7725 |
| **AVG** | **6008** | **5998.6** | **99.84** | **231.4** | **154.45** | **0.054** | **308.76** |

Table 4. Results of AODV routing algorithm

| Network Size | Packet Sent | Packet Received | PDF | EED | Throughput (kbps) | Overhead | Consumed Energy |
|---|---|---|---|---|---|---|---|
| 15 | 4961 | 4520 | 91.11 | 166.907 | 120.13 | 1.479 | 51.0281 |
| 30 | 5419 | 4041 | 89.42 | 158.572 | 110.84 | 3.936 | 123.672 |
| 50 | 4540 | 4066 | 89.56 | 150.093 | 108.84 | 5.342 | 190.0928 |
| 75 | 4119 | 3693 | 89.66 | 170.03 | 100.86 | 8.088 | 294.2147 |
| 100 | 3893 | 3467 | 89.06 | 199.449 | 97.91 | 11.097 | 401.5922 |
| 150 | 3400 | 3017 | 88.74 | 174.209 | 82.3 | 17.413 | 617.8012 |
| **AVG** | **4238.6** | **3800.6** | **89.59** | **169.87** | **103.48** | **7.892** | **279.733** |

Table 5. Results of AOMDV routing algorithm

| Network Size | Packet Sent | Packet Received | PDF | EED | Throughput (kbps) | Overhead | Consumed Energy |
|---|---|---|---|---|---|---|---|
| 15 | 4763 | 4459 | 93.62 | 206.31 | 118.52 | 0.998 | 65.505 |
| 30 | 4768 | 4438 | 93.08 | 172.063 | 118.88 | 2.152 | 135.1011 |
| 50 | 4522 | 4249 | 93.96 | 224.969 | 112.16 | 3.798 | 185.7059 |
| 75 | 4621 | 4314 | 93.36 | 188.46 | 115.39 | 5.611 | 325.1866 |
| 100 | 4173 | 3877 | 92.91 | 188.665 | 103.02 | 8.393 | 401.5922 |
| 150 | 4202 | 3877 | 91.69 | 210.605 | 104.88 | 12.15 | 617.801 |
| **AVG** | **4508.1** | **4198.5** | **93.1** | **198.51** | **112.14** | **5.517** | **288.48** |

Table 6. Results of SARA routing algorithm for 104 nodes

| Simulation Time (s) | Network Size | Packet Sent | Packet Received | PDF | EED | Throughput (kbps) | Consumed Energy |
|---|---|---|---|---|---|---|---|
| 50 | 104 | 2640 | 2605 | 98.67 | 303.269 | 219.5 | 131.753 |
| 100 | 104 | 7263 | 7179 | 98.84 | 243.274 | 301.65 | 247.971 |
| 150 | 104 | 12732 | 12640 | 99.28 | 261.995 | 350.45 | 418.055 |
| 200 | 104 | 14986 | 14888 | 99.25 | 283.553 | 307.7 | 535.815 |
| 250 | 104 | 17711 | 17562 | 99.16 | 339.657 | 291.78 | 709.25 |
| **AVG** | **104** | **11066.4** | **10974.8** | **99.04** | **286.349** | **294.216** | **413.9668** |

Table 7. Results of PAR routing algorithm for 104 nodes

| Simulation Time (s) | Network Size | Packet Sent | Packet Received | PDF | EED | Throughput (kbps) | Consumed Energy |
|---|---|---|---|---|---|---|---|
| 50 | 104 | 2641 | 2591 | 98.56 | 268.517 | 219.95 | 120.792 |
| 100 | 104 | 7980 | 7934 | 99.3 | 168.221 | 341.23 | 257.882 |
| 150 | 104 | 12968 | 12846 | 99.06 | 231.339 | 360.91 | 405.871 |
| 200 | 104 | 15940 | 15940 | 99.28 | 339.26 | 339.26 | 564.782 |
| 250 | 104 | 18385 | 18385 | 99.23 | 269.58 | 306.72 | 695.647 |
| **AVG** | **104** | **11634.2** | **11539.2** | **99.086** | **255.383** | **313.614** | **408.9948** |

## ACKNOWLEDGEMENTS

The author would like to acknowledge the funding support from UGC under the RGNF fellowship scheme (Ref.No.F1-17.1/2012-13/RGNF-2012-13-SC-KAR- 17563/(SAIII/Website) dated February 28,2013, government of India. Thanks also go to the dedicated research group in the area of Computer Networking at the Dept of Computer Science, Mangalore University, Mangalore, India, for many valuable discussions. Lastly but not least the author would like to thank everyone, including the anonymous reviewer.